\documentclass[twocolumn,amsmath,aps,nofootinbib]{revtex4}
       
\usepackage{amssymb}
\usepackage{amsmath}
\usepackage{epsfig}
\usepackage{subfigure}
\usepackage{mathrsfs}

\newcommand{\be}{\begin{equation}}
\newcommand{\ee}{\end{equation}}
\newcommand{\bea}{\begin{eqnarray}}
\newcommand{\eea}{\end{eqnarray}}
\newcommand{\ba}{\begin{eqnarray}}
\newcommand{\ea}{\end{eqnarray}}

\newcommand{\sdel}{\vec{\nabla}}
\renewcommand{\d}{\mathrm{d}}
\newcommand{\p}{\partial}
\newcommand{\lm}{{(\ell m)}}

\newcommand{\T}{{\text{\tiny\it (T)}}}
\newcommand{\TF}{{\text{\tiny\it (TF)}}}
\newcommand{\GI}{{\text{\tiny\it (GI)}}}
\newcommand{\GMG}{{\mathrm{\text{\tiny\it GMG}}}}
\renewcommand{\H}{\mathcal{H}}

\newcommand{\gmgphi}{\phi}
\newcommand{\gmgeta}{\eta}
\newcommand{\gmgpsi}{\varsigma}
\newcommand{\gmgchi}{\chi}
\newcommand{\gmgPi}{\Pi}
\newcommand{\gmgk}{\varphi}
\newcommand{\gmgalpha}{v}
\newcommand{\gmgbeta}{\bar{v}}
\newcommand{\gmggamma}{w}
\newcommand{\gmgomega}{\Delta}

\begin{document}

\title{Perturbation Theory in Lema\^itre-Tolman-Bondi Cosmology}

\author{Chris Clarkson$^1$, Timothy Clifton$^2$ and Sean February$^1$}

\affiliation{$^1$ Cosmology \& Gravity Group, Department of
  Mathematics and Applied Mathematics, University of Cape Town,
  Rondebosch 7701, Cape Town, South Africa\\ $^2$ Department of
  Astrophysics, University of Oxford, Oxford, OX1 3RH, UK}

\email{tclifton@astro.ox.ac.uk}

% ----------------------- ABSTRACT -------------------------

\begin{abstract}

The Lema\^itre-Tolman-Bondi solution has received much attention as a
possible alternative to Dark Energy, as it is able to account for the
apparent acceleration of the Universe without any exotic matter
content.  However, in order to make rigorous comparisons between these
models and cosmological observations, such as the integrated
Sachs-Wolfe effect, baryon acoustic oscillations and the observed
matter power spectrum, it is absolutely necessary to have a proper
understanding of the linear perturbation theory about them.  Here we
present this theory in a fully general, and gauge-invariant form.  It
is shown that scalar, vector and tensor perturbations interact, and
that the natural gauge invariant variables in Lema\^itre-Tolman-Bondi
cosmology do not correspond straightforwardly to the usual Bardeen
variables, in the limit of spatial homogeneity. We therefore construct
new variables that reduce to pure scalar, vector and tensor modes in
this limit. 

\end{abstract}

\maketitle

%\tableofcontents

% ---------------------- INTRO -------------------------------------------

%\vspace{-10pt}
\section{Introduction}
%\vspace{-10pt}

The concordance model of cosmology is a result of assuming that the
Universe is approximately the same at all points in space, and that
at every point it appears isotropic. The observations we
make locally can then be extrapolated to the Universe at large, and
the resulting Friedmann-Lema\^itre-Robertson-Walker
(FLRW) model can be used to interpret all cosmological observations.
The FLRW models are, of course, very well understood, and have
proved to be highly successful in very many different capacities.
Unfortunately, despite its aesthetic appeal and mathematical
simplicity, the adoption of the FLRW cosmology has led to the undesirable requirement
that the Universe must be filled with a smoothly distributed and
gravitationally repulsive exotic substance, known as Dark Energy.  The
conceptual problems associated with Dark Energy are well known, and
are distressing enough that it now seems worth investigating
alternative, less appealing models.

One such alternative is the spherically symmetric, dust
dominated cosmological solution of Lema\^{i}tre, Tolman and Bondi
\cite{LTB}.  This model is isotropic about a single point in space, in
agreement with local galaxy surveys and observations of the Cosmic Microwave
Background, but is anisotropic everywhere else~\cite{MM}.  As such, we are
required to live very near the centre of symmetry, indicating a break
with the Copernican Principle that {\it ``the Earth is not in a
central, specially favoured position''} \cite{cop}.  The
Lema\^{i}tre-Tolman-Bondi (LTB) model allows the Hubble rate to vary with
radial distance from the centre of symmetry, as well as with time, and
so allows for the possibility of apparent acceleration (if observations
within them are interpreted in an FLRW frame-work), while no part of them
need undergo any accelerating expansion.

One may then try and argue on aesthetic grounds, or on the grounds of
presumed early Universe physics, which of these models is more
likely.  Clearly there are good reasons to disfavour either, depending
on which one abhors most: the Earth being in a special position in the Universe,
or the existence of Dark Energy, which requires us to live at a
special time when the acceleration is just starting to kick in.
However, a more objective (and scientific) approach to 
the problem would be to try and distinguish between the two
models with direct observations. This has been
attempted a number of different times in the literature already
\cite{voids}. However, in order to make comparisons to different
types of astrophysical data it is absolutely necessary to have a
proper understanding of how structures form and behave in these
models, and this means understanding their perturbation theory.  An
alternative approach is to try to test the Copernican
principle directly~\cite{copp}, but this is very difficult to achieve at present
through direct tests.   

For observations such as the integrated Sachs-Wolfe effect,
Rees-Sciama effect, or matter power spectrum it is obvious that a
proper interpretation of the data is dependent on understanding the
relevant perturbation theory.  However, other observations, that do
not necessarily require perturbation theory in FLRW, will also require
a knowledge of perturbation theory in the LTB case.  An example is the
use of baryon acoustic oscillations as a probe of cosmology.  In FLRW
it suffices to consider comoving points in an unperturbed fluid, to
see how length scales at recombination are evolved onto distance
measures on our past light cone.  In an LTB universe, however, the
rate of growth of structure can be different at different places, and
the peculiar velocity of perturbations need not cancel when averaged
over suitable scales (in fact, one may naively expect otherwise).
Likewise, when considering light reflecting off a galaxy cluster in
FLRW it may suffice to consider that on average the motions of galaxy
clusters will cancel out, so that approximating galaxy clusters as
being comoving with the background fluid may be sufficient to get
accurate results.  In an LTB universe this is unlikely to be true, as
gradients in gravitational potentials will develop due to different
growth rates at different positions.

As well as the requirement of understanding linear perturbation theory
in order to accurately account for already established probes of cosmology, there
is also the possibility of finding new observational phenomena that
could be used to distinguish the LTB and FLRW cosmologies, or
constrain the type of LTB models that are observationally viable.  An
example of this is the potential coupling of scalar gravitational
potentials to tensor gravitational waves.  In an FLRW background (to
linear order) the decomposition theorem ensures that no such coupling
occurs. On a homogeneous background scalar, vector and tensor modes
then evolve independently.   In an inhomogeneous LTB background, however, the
decomposition theorem will only be valid on the two dimensional surfaces of spherical
symmetry.  In the three dimensional space there will be no such theorem,
and in general one should expect coupling to occur between what we conventionally think of as `scalar', `vector'  and `tensor' modes.
Though complicated, such effects could potentially be used as probes of the geometry of the universe.

In section \ref{sec2} we introduce the formalism required for
perturbation theory about a spherically symmetric background.  In
section \ref{sec3a} and \ref{sec3b} we present the perturbation equations, and in
section \ref{sec4a} and \ref{sec4b} we discuss their solutions and how they relate to
perturbation theory in FLRW.  Finally, in section \ref{sec5} we
conclude.  The Appendices contain information about gauge invariant
quantities for the metric perturbation, and the perturbation equations
for some special cases.

We use units in which $c=G=1$, throughout.

\section{Perturbation theory on a spherically symmetric background}
\label{sec2}

Perturbations on a spherically symmetric space-time have been
considered a number of times in the literature, mostly in the context
of modelling static and stationary stars \cite{stars}.  A time
dependent formalism was first developed by Seidel \cite{seidel}, and
later specialised to a space-time containing a perfect fluid by
Gundlach and Mart\'{i}n-Garc\'{i}a (GMG) \cite{gmg}.
Perturbations of self similar models have been
investigated in~\cite{ss}, and have the advantage that all PDEs reduce
to ODEs in the system of perturbation equations. A limited class of
perturbations in LTB cosmology have also been studied by Zibin
\cite{zibin}. 

Different formalisms exist in the literature for constructing both the system of perturbation
equations, and the gauge-invariant variables. Gerlach and
Sengupta~\cite{gersen} (GS) developed a formalism based on a 2+2
covariant split of the metric and energy-momentum tensor.  This turns the
field equations into a system of second-order PDEs, and was made explicit by GMG, who found a closed set of master equations
describing all perturbations. Their approach was used in the study of
self-similar LTB perturbations performed in~\cite{ss}. Alternatively, a
covariant 1+1+2 formalism has been developed in~\cite{1+1+2}, which builds
on the covariant 1+3 formalism that has been usefully applied in
cosmology~\cite{1+3}.  Here the Bianchi
and Ricci identities, plus the Ricci rotation coefficients for the
semi-tetrad introduced, are covariantly split into a system of first-order differential
equations. This formalism has not yet been reduced to a tractable set
of master equations for a general space-time, but has been used by~\cite{zibin} to study LTB perturbations.   
Here we apply the GS formalism developed by GMG to the case of an LTB
space-time.  This results in a simple set of coupled  second-order
PDEs that describe general perturbations to the space-time, in a gauge
invariant way.

\subsection{The LTB background}

The unperturbed LTB line-element can be written
\be
\d s^2 = -\d t^2 + \frac{a_{\parallel}^2(t,r)}{(1-\kappa r^2)} \d r^2 + a_{\perp}^2(t,r)
r^2 \d \Omega^2,
\ee
where $a_{\parallel}=(r a_{\perp})_{,r}$ and $\kappa=\kappa(r)$ is a
free function of $r$.  The
FLRW scale factor, $a$, has been replaced here by two new scale
factors, $a_{\parallel}$ and $a_{\perp}$, describing  expansion
parallel and perpendicular to the radial direction, respectively.  For
future use we will define the radial and azimuthal Hubble rates to be
\be
H_{\parallel} \equiv \frac{\dot{a}_{\parallel}}{a_{\parallel}}
~~~~\text{and}~~~~
H_{\perp} \equiv \frac{\dot{a}_{\perp}}{a_{\perp}},
\ee
where an over-dot denotes partial differentiation with respect to
$t$.  The analogue of the Friedmann equation in this space-time is
then given by
\be
\label{H}
H_{\perp}^2 = \frac{M}{a_{\perp}^3}-\frac{\kappa}{a_{\perp}^2},
\ee
where $M=M(r)$ is another free function of $r$, and the locally measured energy density is 
\be
8 \pi \rho = \frac{(M r^3)_{,r}}{a_{\parallel} a_{\perp}^2 r^2},
\ee
which obeys the conservation equation
\be
\dot{\rho}+ (2 H_{\perp}+H_{\parallel}) \rho =0.
\ee
The acceleration equations in the perpendicular and parallel directions are
\be
\frac{\ddot{a}_{\perp}}{a_{\perp}} = -\frac{M}{2 a_{\perp}^3}
~~~~~\text{and}~~~~
\frac{\ddot{a}_{\parallel}}{a_{\parallel}} = -4 \pi \rho +\frac{M}{a_{\perp}^3}.
\ee
For what follows it will also be useful to define the radial derivative
\be
X^{\prime} \equiv \frac{\sqrt{1-\kappa r^2}}{a_{\parallel}} X_{,r}.
\ee
This derivative does not commute with the time derivative, but instead obeys 
\be
(\dot X)'-{(X')}{\dot{}}= H_\parallel X'.
\ee
We also define the curvature function
\be
W \equiv \frac{\sqrt{1-\kappa r^2}}{a_{\perp} r}.
\ee
The following relations are then obeyed
\ba
H_{\perp}^{\prime} &=& W (H_{\parallel}-H_{\perp}),\\
\dot{W} &=& - H_{\perp} W,\\
W' &=& -W^2 - 4 \pi \rho+H_\perp H_{\parallel} + \frac{M}{2 r^3}.
\ea
In the perturbation equations that follow we will choose to eliminate $\kappa$
in favour of $W$, so that the equations take their simplest form.

\subsection{Harmonic functions}

A natural way to split this space-time is in a 2+2 decomposition,
so that the space-time manifold becomes $M^4=M^2 \times S^2$,
where $S^2$ indicates the 2 dimensional spherically symmetric surfaces.  We will use lower case Latin
indices $a,b,c,\ldots$ to denote coordinates in $S^2$, upper case
Latin indices $A,B,C,\ldots$ to denote
coordinates in $M^2$, and Greek indices $\mu,\nu,\xi,\ldots$ to denote coordinates that run over all 4 space-time dimensions.

In FLRW cosmology, any perturbation can be split into scalar, vector and
tensor (SVT) modes that decouple from each other, and so evolve
independently (to first order).  This classification is based on how they
transform on the homogeneous and isotropic spatial hyper-surfaces, and
is essentially just a generalisation of Helmholtz's
theorem~\cite{stewart}.

Such a split cannot usefully be performed in the same way here, as the
background is no longer spatially homogeneous, and modes written in this way would couple together (as we shall see).  However, one can
perform an analogous classification based on how the
perturbations transform on the surfaces of spherical symmetry.  This
 results in a decoupling of the perturbations into two independent
 modes, called `polar' (or even) and `axial' (or odd), which are analogous, but
 not equivalent, to scalar and vector modes in
 FLRW. Unlike the FLRW case, however, there is no further decomposition into
 tensor modes as no non-trivial symmetric, transverse and
 trace-free rank 2 tensors can exist on $S^2$.  Therefore, only two distinct sectors exist. 
Scalars (rank 0 tensors) on $S^2$ can then be expanded as a sum of polar modes, and
higher rank tensors on $S^2$ can be expanded in sums over both the polar and
axial modes. Only a scalar can contain spherical perturbations (given
by $\ell=0$, defined below), and only scalars and vectors (i.e.,
tensors of rank 0 and 1) can contain a dipole term ($\ell=1$). Higher
multipoles can be present in all tensors. 

An appropriate family of basis functions for this split
are tensor spherical harmonics.  These are derived from the usual
spherical harmonic functions, $Y^\lm(x^a)$, that obey
\be
\bar{\nabla}^2 Y^\lm = - \ell  (\ell +1) Y^\lm,
\ee
where the Natural number, $\ell $, gives the angular scale of the perturbation.  
The Laplacian, $\bar{\nabla}^2$ here, is on the surface of spherical symmetry, and
is given by $\bar{\nabla}^2 \phi = \gamma^{ab} \phi_{:ab}$,
where the colon subscript indicates a covariant derivative with
respect to the metric on the unit sphere, $\gamma_{ab}$.  Scalar
perturbations on $S^2$ can then be written with their angular
dependence given in terms of the solutions to this equation, and expanded as 
\be
\phi(x^A,x^a)=\sum_{\ell=0}^\infty\sum_{m=-\ell}^{m=\ell} \phi^\lm(x^A)Y^\lm(x^a).
\ee

It is now possible to construct a basis for all higher rank tensors
from $Y^\lm$, its covariant derivatives, and the contractions of those derivatives with the
fundamental antisymmetric tensor, $\epsilon_{ab}$.
Modes that can be described without requiring $\epsilon_{ab}$ are called polar, while
those that require $\epsilon_{ab}$ are called axial. 

We can now form harmonic functions for higher rank tensor
perturbations with polar degrees of freedom by first defining the vector, for $\ell\geq1$:
\be
Y^{(\ell m)}_a \equiv Y^{(\ell m)}_{:a}.
\ee
 We define also the trace-less tensor, for $\ell\geq2$:
\be
Y^{(\ell m)}_{ab} \equiv Y^{(\ell m)}_{:ab}+\frac{\ell  (\ell +1)}{2} Y^{(\ell m)} \gamma_{ab}.
\ee
Taking divergences of $Y^{(\ell m)}_a$ and $Y^{(\ell m)}_{ab}$ reduces these expressions to equations involving $Y^{(\ell m)}$. 

For axial perturbations on $S^2$ we define a divergence-free vector
harmonic, for $\ell\geq1$,
\be
\bar Y^\lm_a\equiv\epsilon_a^{~b}Y^\lm_{:b}.
\ee
We can then construct a symmetric and trace-free rank-2 axial harmonic function by defining, $\ell\geq2$,
\be
\bar Y^\lm_{ab}\equiv 2 \bar
Y^\lm_{(a:b)}=-2 {\epsilon^d}_{(a} Y^\lm_{: b) d},
\ee
where round brackets around indices denote symmetrisation, as usual.
Taking covariant derivatives, this
tensor harmonic can be reduced to an expression involving $\bar Y^\lm_a$. 
 
For both parities, the `vector harmonics' obey
\be
\bar{\nabla}^2 Y^\lm_a = [1-\ell (\ell +1)]  Y^\lm_a
\ee
(similarly for $\bar Y^\lm_a$), and the `tensor harmonics' obey
\be
\bar{\nabla}^2 Y^\lm_{a b} = [4-\ell (\ell +1)]  Y^\lm_{a b}.
\ee

Because the vector harmonics are orthogonal for each $\ell$, any
rank-1 tensor perturbation can now be expanded as 
\ba
\phi_a(x^A,x^a)=\sum_{\ell=1}^\infty\sum_{m=-\ell}^{m=\ell}
&&\phi^\lm(x^A)Y_a^\lm(x^a)\nonumber\\+ &&\bar\phi^\lm(x^A)\bar Y_a^\lm(x^a),
\ea
where $\phi^\lm$ and $\bar\phi^\lm$ are independent for each $\lm$,
and are given by
\be
\phi^\lm=-\frac{1}{\ell(\ell+1)}\int\d\Omega\,( {\phi^a}_{:a}) Y_\lm^*
\ee
and
\be
\bar\phi^\lm=-\frac{1}{\ell(\ell+1)}\int\d\Omega\,(\epsilon^{ab} \phi_{b:a}) Y_\lm^*.
\ee
There are no $\ell=0$ vector dof because $\ell=0$ describes spherical modes.

Finally, let us consider trace-less rank 2 tensor perturbations on $S^2$.
A suitable orthogonal basis in this case, for $\ell\geq2$, is
given by the two harmonic functions $Y^\lm_{ab}$ and $\bar Y^\lm_{ab}$,
defined above.  Any rank 2 trace-free tensor perturbation can then be written
\ba
\phi_{ab}(x^A,x^a)=\sum_{\ell=2}^\infty\sum_{m=-\ell}^{m=\ell}
&&\phi^\lm(x^A)Y_{ab}^\lm(x^a)\nonumber\\+&&\bar\phi^\lm(x^A)\bar
Y_{ab}^\lm(x^a),
\ea
where
\ba
\phi^\lm&=&2\frac{(\ell-2)!}{(\ell+2)!}\int\d\Omega\,({\phi^{ab}}_{:ba}) Y_\lm^*,\nonumber\\
\bar\phi^\lm&=&2\frac{(\ell-2)!}{(\ell+2)!}\int\d\Omega\,(\epsilon_{a}^{~c}
{\phi^{ab}}_{:bc}) Y_\lm^*.
\ea

As
our background is spherically symmetric, all perturbations with
different $(\ell m)$ decouple from each other.  This is analogous to
the decoupling of Fourier modes in FLRW.  We shall therefore drop the
$\lm$ labels on all quantities that follow. Where a function of $x^A$
is multiplied by a harmonic function, a sum over $\ell$ and $m$ is implied.

\subsection{Metric perturbations}

Gauge invariant
variables for general perturbations of a spherically symmetric background, with
arbitrary matter content, have been formulated by Gerlach and Sengupta (GS) \cite{gersen}.
We reiterate their results, relevant to the present study, in
Appendix \ref{A}.  We then review the perfect fluid formalism of GMG,
specialised to a dust filled universe, in Appendix \ref{B}.  These
studies show that there is a preferred gauge, known as the
Regge-Wheeler (RW) gauge~\cite{RW}, in which the perturbation variables are
equal to gauge invariant quantities (comparable to the longitudinal, or conformal Newtonian gauge, of FLRW).  The RW
gauge is the choice that any off-diagonal polar modes in the metric that have an
angular index are zero, and that in the axial modes there is no
perturbation to the $g_{ab}$ components (i.e. the $S^2$
components).  Calculations performed in the RW gauge are
then equivalent to those performed when considering the gauge invariant variables of
\cite{gmg} and \cite{gersen}.

The general form of polar perturbations to the metric can now be
written, in RW gauge, as\footnote{We have changed notation from GMG to
 avoid potential confusion with notation that is often used in
 cosmology. In particular we have replaced: $k\mapsto\gmgk,
 \chi\mapsto\gmgchi, \psi\mapsto\gmgpsi,\eta\mapsto\gmgeta, \alpha\mapsto\gmgalpha,
 \beta\mapsto\gmgbeta, \omega\mapsto\gmgomega, \gamma\mapsto\gmggamma$. } 
\begin{align}
\label{gpolar}
\d s^2 = &-\left[1+(2 \gmgeta-\gmgchi-\gmgk) Y \right] \d t^2 -\frac{2
a_{\parallel} \gmgpsi Y}{\sqrt{1- \kappa r^2}} \d t \d r\\
&+\left[1+(\gmgchi+\gmgk) Y \right] \frac{a_{\parallel}^2
\d r^2}{(1-\kappa r^2)} +a_{\perp}^2 r^2 (1+ \gmgk Y) \d\Omega^2, \nonumber
\end{align}
where $\gmgeta(t,r)$, $\gmgchi(t,r)$, $\gmgk(t,r)$ and $\gmgpsi(t,r)$ are equal to the gauge
invariant quantities of GS and GMG, as shown in Appendices \ref{A} and
\ref{B}.  The general form of polar matter perturbations in this gauge
are parameterised by
\bea
\label{upolar}
u_{\mu} &=& \left[\hat{u}_A+ \left( \gmggamma \hat{n}_A +\frac{1}{2} h_{AB}
\hat{u}^B \right) Y, \gmgalpha Y_{a} \right]\\
\label{rhopolar}
\rho &=& \rho^{LTB} (1+\gmgomega Y),
\eea
where $\gmgalpha$, $\gmggamma$ and $\gmgomega$ are equal to the gauge
invariant quantities of GS and GMG, again, as shown in Appendices
\ref{A} and \ref{B}.  The vectors $\hat{u}^A$ and $\hat{n}^A$ indicate
background unit vectors in the time-like and space-like radial directions,
respectively.  They are given by
\be
\hat{u}^A = (1,0)
\qquad
\text{and}
\quad
\hat{n}^A = \left( 0, \frac{\sqrt{1- \kappa r^2}}{a_{\parallel}} \right).
\ee
The $h_{AB}$ in (\ref{upolar}) correspond to the linear perturbations
to $g_{AB}$, as shown in (\ref{gpolar}), and are included to ensure
the normalisation $u^{\mu} u_{\mu}=-1$.

The general form of axial perturbations to the metric, in RW gauge, are
\begin{align}
\label{gaxial}
\d s^2 = &-\d t^2  +\frac{a_{\parallel}^2}{(1-\kappa r^2)}\d r^2
+a_{\perp}^2 r^2 \d\Omega^2\\ &+ 2 k_A \bar Y_b \d x^A \d x^b,
\end{align}
where $k_A$ is equal to the gauge invariant perturbation of GS, from
Appendix \ref{A}.  Following GS, we also define for later convenience the new variable
\be
\label{Pi}
\gmgPi \equiv \epsilon^{AB} \left( \frac{k_A}{a_{\perp}^2 r^2} \right)_{\vert B},
\ee
where the fundamental anti-symmetric tensor
$\epsilon_{AB}=n_Au_B-u_An_B$ and a pipe denotes the covariant
derivative on $M^2$.
The only axial perturbation that can then occur to the matter content is
a perturbation to the matter four velocity, such that
\be
\label{uaxial}
u_{\mu} = (\hat{u}_A, \gmgbeta \bar Y_a),
\ee
where $\hat{u}^A$ is defined as in the polar case, and $\gmgbeta$
equals one of the gauge invariant variables of GMG, in Appendix \ref{B}.

\subsubsection{`Scalar-Vector-Tensor' variables}
\label{newvars}

The GI variables defined above give a very concise set of governing
equations (see below). However, in the FLRW limit we shall see that
they mix up the normal SVT modes in a complicated way~-- for example,
the variable $\gmgk$ contains all three types of perturbations. By
taking combinations of the variables defined in the perturbed metric
we may form variables which do reduce to scalars, vectors, or tensors
in the FLRW limit, and so may be used to identify generalised SVT
modes. Such variables are: 

\noindent\emph{Tensors}
 \ba
\label{te1}
 \text{polar:}~~~ \hat\gmgchi&=&\chi\\ \label{te2}
\text{axial:}~~~ \hat{\Upsilon}&\equiv&\gmgPi''+6W\gmgPi'+\left(8W^2-\frac{\ell(\ell+1)+2}{a_\perp^2r^2}\right)\gmgPi\nonumber\\&&
+16\pi\frac{(\rho\gmgbeta)'}{a_\perp^2r^2},
\ea

\noindent\emph{Vectors}
\ba
\label{ve1}
 \text{polar:}~~~\hat{\xi}&\equiv&\frac{3a_\perp}{2W}\Bigg[\frac{1}{r^3}\left(r^2\dot\gmgchi\right)'+\left(\frac{\gmgpsi}{r}\right)''+2W\left(\frac{\gmgpsi}{r}\right)'
\nonumber\\ \label{ve2} &&~~~ -\left(\frac{\ell(\ell+1)-3}{a_\perp^2r^2}+3W^2\right)\frac{\gmgpsi}{r}\Bigg] \\
\text{axial:}~~~ \hat{\gmgbeta}&=&\gmgbeta,
\ea

\noindent\emph{Scalars}
\ba
\hat{\zeta}&\equiv&{\hat\lambda}''+2W{\hat\lambda}'-\frac{\ell(\ell+1)}{a_\perp^2
 r^2}{\hat\lambda} 
\nonumber\\ \label{sc2} &&~~~~~~~~~+rW\hat\xi'+r\left(3W^2-\frac{1}{a_\perp^2r^2}\right)\hat\xi,
\ea

where
\be
\hat{\lambda} \label{sc1} \equiv 8\pi \rho a_\perp \left[H_\perp^{-1}\gmgomega-3\gmgalpha\right].
\ee

We shall justify these variables in Sec.~\ref{NV}.

\section{Master equations }

The equations of motion governing the dynamics may be reduced to a
coupled system of evolution equations and constraints. For the polar
sector the equations come in the form of three coupled PDEs for
$\gmgchi,\gmgk$ and $\gmgpsi$, with all other non-trivial gauge invariant variables
determined from the solution of this system. For the axial modes, the
dynamics are determined by a much simpler system of equations for the
variables $\gmgPi$ and $\gmgbeta$.

%\begin{widetext}

\subsection{Polar perturbation equations for $\ell\geq2$}
\label{sec3a}

Substituting the polar perturbed metric tensor (\ref{gpolar}), and the
perturbed matter quantities, (\ref{upolar}) and (\ref{rhopolar}), into
the field equations results in a system of three coupled master
evolution equations for the three variables $\gmgchi,\gmgk$ and $\gmgpsi$.
The remaining variables associated with the fluid can then be
determined directly from the solution to this system. 

The three evolution equations are
\begin{widetext}
\bea
\label{evo1}
&& -\ddot{\gmgchi} + \gmgchi^{\prime \prime} -3 H_{\parallel}\dot{\gmgchi} -2 W \gmgchi^{\prime}  + \left[ 16 \pi \rho -\frac{6M}{a_{\perp}^3}-4 H_{\perp} (H_{\parallel}-H_{\perp})
-\frac{(\ell -1)(\ell +2)}{a_{\perp}^2 r^2}\right] \gmgchi 
\\ \nonumber
&&~~~~=- 2 (H_{\parallel} -H_{\perp})
\gmgpsi^{\prime}-2  \left[
H_{\parallel}^{\prime}-2(H_{\parallel}-H_{\perp}) W \right] \gmgpsi
 +4 (H_{\parallel} -H_{\perp}) \dot{\gmgk} -2 \left[ 8 \pi \rho -\frac{3M}{a_{\perp}^3}-2 H_{\perp} (H_{\parallel}-H_{\perp})
\right]  \gmgk , 
\eea
and
\be
\label{evo2}
\ddot{\gmgk}+ 4 H_{\perp} \dot{\gmgk}-2 \left( \frac{1}{a_{\perp}^2
  r^2}-W^2 \right) \gmgk =  -H_{\perp} \dot{\gmgchi}+ W
  \gmgchi^{\prime} -\left[ 2 W^2 -\frac{\ell (\ell +1)+2}{2 
  a_{\perp}^2 r^2} \right] \gmgchi +2 W
(H_{\parallel} -H_{\perp}) \gmgpsi ,
\ee
and
\be
\label{evo3}
\dot{\gmgpsi} + 2 H_{\parallel} \gmgpsi = -\gmgchi^{\prime},
\ee
together with the constraint 
\be
\gmgeta=0.
\ee  
These four equations are
the LTB version of Equations (GMG87), (GMG88), (GMG89) and (GMG82) from \cite{gmg}. 

The three constraint equations can then be written as
\bea
\label{con1}
8 \pi \rho \gmggamma &=& ( \dot{\gmgk} )^{\prime}- (H_{\parallel} -2 H_{\perp})
\gmgk^{\prime} - W \dot{\gmgchi}
+H_{\perp} \gmgchi^{\prime} 
+ \left[ \frac{\ell  (\ell +1)+2}{2 a_{\perp}^2 r^2} +H_{\perp}^2 +2 H_{\perp} H_{\parallel} - W^2 -4 \pi \rho \right] \gmgpsi,\\
\label{con2}
8 \pi \rho \gmgomega&=& -\gmgk^{\prime \prime} - 2 W \gmgk^{\prime}+(H_{\parallel}+2 H_{\perp}) \dot{\gmgk}+W
\gmgchi^{\prime} + H_{\perp} \dot{\gmgchi} + \left[ \frac{\ell (\ell +1)}{a_{\perp}^2r^2} +2 H_{\perp}^2 +4 H_{\parallel} H_{\perp} -8 \pi \rho \right]
(\gmgchi +\gmgk) \\ \nonumber && - \frac{(\ell -1)(\ell +2)}{2 a_{\perp}^2r^2} \gmgchi  +2 H_{\perp} \gmgpsi^{\prime}+2 (H_{\parallel}+H_{\perp}) W
\gmgpsi ,\\
\label{con3}
8 \pi \rho \gmgalpha &=&
\dot{\gmgk} +\frac{\dot{\gmgchi}}{2}  + H_{\parallel} (\gmgchi+\gmgk)+ \frac{\gmgpsi^{\prime}}{2}.
\ea
These are the LTB versions of the equations (GMG93), (GMG94) and
(GMG95) from \cite{gmg}.  It is also useful to consider the
evolution equations that result from differentiating (\ref{con1}),
(\ref{con2}) and (\ref{con3}).  These take the particularly simple
form
\ba
\label{mat1}
\dot{\gmgalpha}&=& \frac{\gmgchi}{2} +\frac{\gmgk}{2},\\
\label{mat2}
\dot{\gmggamma} &=& \frac{\gmgk^{\prime}}{2} - H_{\parallel} \gmggamma -
\frac{H_{\parallel}}{2} \gmgpsi,
\\
\label{mat3}
\dot{\gmgomega}+\left( \gmggamma +\frac{\gmgpsi}{2}
\right)^{\prime} &=& -\frac{\dot{\gmgchi}}{2} -\frac{3
\dot{\gmgk}}{2}
+\frac{\ell  (\ell +1)}{a_{\perp}^2r^2} \gmgalpha 
 - \frac{\rho^{\prime}}{\rho} \left(
\gmggamma +\frac{\gmgpsi}{2} \right) 
-2 W \left( \gmggamma +\frac{\gmgpsi}{2} \right),
\eea
\end{widetext}
and are the LTB versions of (GMG96), (GMG97) and (GMG98). There are
then six coupled equations for the six variables
$\gmgchi$, $\gmgk$, $\gmgpsi$, $\gmggamma$, $\gmgomega$ and $\gmgalpha$.

Equations (\ref{evo1})-(\ref{mat3}) are clearly more complicated than their FLRW
counterparts, but the fact that they can be written as concisely as
they are above is quite remarkable.

For very large angle fluctuations, with $\ell =0$ or $1$, the field
equations no longer give $\gmgeta=0$, which has been used to simplify the equations in this section.  Instead,  there are
additional gauge freedoms that can be used to simplify
(\ref{evo1})-(\ref{mat3}).  These are discussed in Appendix~\ref{C}.

\subsection{Axial perturbation equations}
\label{sec3b}

The axial perturbation equations take a simpler form than their polar
counterparts.  Because we are considering dust dominated
cosmologies, the field equations give us $\dot{\gmgbeta}=0$, and so
$\gmgbeta=\gmgbeta (r)$ and must be set by initial conditions.  For
$\ell \geq2$ the metric perturbation $\gmgPi$ can be shown to obey the
wave equation
\bea
\label{axi1}
&& - \ddot{\Pi} + \Pi^{\prime \prime} - (6 H_{\perp}
+H_{\parallel}) \dot{\Pi}+ 6 W \Pi^{\prime} \\ \nonumber &&~~~~- \left[ 16 \pi \rho  +
\frac{(\ell -2)(\ell +3)}{a_{\perp}^2r^2} \right] \Pi  = -16 \pi \frac{(\rho
\gmgbeta)^{\prime}}{a_{\perp}^2r^2},
\eea
which is the LTB version of (GMG69).  This is the only dynamical
equation that needs solving, for the single variable
$\gmgPi$.  Once this equation is solved it gives the axial metric
perturbations, $k_A$, via
\be\label{k-axial}
(\ell -1) (\ell +2) k_A = 16 \pi\rho  a_{\perp}^2 r^2 \gmgbeta u_A-
\epsilon_{AB} (a_{\perp}^4r^4\gmgPi)^{\vert B}.
\ee
Clearly $\ell =1$ is a special case.  For large angle perturbations of
this type we have $(a_{\perp}^4r^4 \gmgPi \dot{)}=0$ and
$(a_{\perp}^4r^4 \gmgPi )^{\prime} = -16 \pi a_{\perp}^2 r^2 \rho
\gmgbeta$.  The solution for $\gmgPi$ is then
\be
\Pi = -\frac{2}{a_{\perp}^4r^4} \int \frac{\gmgbeta (M r^3)_{,r}}{\sqrt{1-\kappa r^2}} dr.
\ee
The metric perturbations $k_A$ must then be obtained from inverting
Eq. (\ref{Pi}).  This will involve an extra degree of freedom in $k_A$
that means it cannot be determined uniquely.

There are no axial perturbations with $\ell =0$.

\subsection{Structure of the equations and solutions}
\label{sec4a}

It can be seen from the master equations, (\ref{evo1})-(\ref{mat3}),  that $\gmgchi$ and $\gmgPi$
contain gravitational wave degrees of freedom, as their leading derivatives are of the form
$-(\ddot{~~})+(~)''$ (i.e. a wave equation with a characteristic
speed of unity). It can also be seen
that $\gmgpsi$ and $\gmgbeta$ look rather like vectors in FLRW
cosmology, while $\gmgk$ appears to govern density perturbations like the
Newtonian potential. Unfortunately, this roughly analogous behaviour
does not hold completely, as we will show below when we consider the FLRW
limit. What is clear, however, is that there exist complicated
couplings between the variables in this system.  We can therefore no longer
expect gravitational waves to decouple from density perturbations, as the
varying curvature of the background serves to couple these
perturbations intimately.

The solutions to the perturbation equations in sections \ref{sec3a}
and \ref{sec3b} will, in general, need to be found numerically.
However, if we consider $\ell \geq2$ and neglect the contribution of the gravitational waves
described by $\gmgchi$, we can make progress.  
With $\gmgeta = \gmgchi=0$, Equation (\ref{evo3}) can be integrated to find
\be
\nonumber
\gmgpsi \propto \frac{1}{a_{\parallel}^2}.
\ee
Clearly, $\gmgpsi \rightarrow 0$ as the Universe expands, and
$a_{\parallel} \rightarrow \infty$.  Assuming $\gmgchi$ to be negligible
can then be seen to lead to a negligible $\gmgpsi$ in the late Universe.  With
$\gmgchi=\gmgpsi=0$, Equation (\ref{evo2}) becomes
\be
\label{evo2b}
\ddot{\gmgk} + 4 H_{\perp} \dot{\gmgk} - \frac{2 \kappa}{a_{\perp}^2} \gmgk =0.
\ee
This equation can be solved parametrically, together with the analogue
of the Friedmann equation, (\ref{H}).  There are three solutions, depending on the sign of $\kappa$.

For $\kappa <0$ we find the solution
\bea
a_{\perp} &=& \frac{M (1-\cosh 2 \Theta )}{2 \kappa}\\
\gmgk &=& \frac{\cosh \Theta }{\sinh^5 \Theta } [ c_1 +c_2 (
\sinh 2 \Theta -6 \Theta \\ \nonumber && \qquad \qquad \qquad \qquad
\qquad +4 \tanh \Theta ) ]\\
t-t_0 &=&  \frac{M (\sinh 2 \Theta -2 \Theta)}{2 (-\kappa)^{3/2}},
\eea
where $c_1=c_1(r)$, $c_2=c_2(r)$ and $t_0=t_0(r)$ are free functions
that must be specified as part of the initial conditions.  The
function $t_0(r)$ is the `bang time'.

For $\kappa>0$ we find
\bea
a_{\perp} &=& \frac{M (1-\cos 2 \Theta )}{2 \kappa}\\
\gmgk &=& \frac{\cos \Theta }{\sin^5 \Theta } [ c_1 +c_2 (
\sin 2 \Theta -6 \Theta  \\ \nonumber && \qquad \qquad \qquad \qquad
\qquad +4 \tan \Theta ) ]\\
t-t_0 &=&  \frac{M (2 \Theta - \sin 2 \Theta)}{2 (\kappa)^{3/2}}.
\eea
Again, $c_1$, $c_2$ and $t_0$ are free functions of $r$.

Finally, for $\kappa=0$, a parametric solution is not required, and we find 
\bea
a_{\perp} &=& \frac{(18M)^{1/3} (t-t_0)^{2/3}}{2}\\
\gmgk &=& \frac{c_1}{(t-t_0)^{5/3}} +c_2.
\eea
Once more, with $c_1$, $c_2$ and $t_0$ as free functions of $r$.

In the $\kappa \neq 0$ cases, $\Theta$ is a monotonically
increasing function of $t$, and in all three cases $c_1$ corresponds
to a decaying mode, and $c_2$ to a growing mode.  In the cases with
$\kappa \leq 0$ the `growing mode' is itself either a constant
(for $\kappa=0$) or a decreasing function of $t$ (for $\kappa <0$),
and in those cases we mean growing with respect to the other mode,
which is decaying faster.

For $\kappa =0$ the power-law forms of $a_{\perp}(t)$ and $\gmgk$ mean that the asymptotic
form of both the growing and decaying modes is clear.  It can also be seen
for $\kappa \neq 0$ that both modes, together with $a_{\perp}$, behave as if
$\kappa =0$ in the limit $t\rightarrow t_0$ (as expected, due to their
dynamical evolution being dominated by $\rho$, and not $\kappa$, at
early times).  In the case of $\kappa <0$ the growing mode can be seen
to decay as $\sim 1/(t-t_0)$ when $t \rightarrow \infty$, and the scale
factor, $a_{\perp}$, behaves like an open Milne universe.  For $\kappa >
0$ the scale factor initially grows, reaches a maximum of expansion,
and then collapses to zero in finite time, $t$.  As this future
singularity is approached, both growing and decaying modes of $\gmgk$
diverge to infinity.

These solutions may serve as the basis for calculating the matter
power spectrum. However, they will only remain so as long as the
evolution equation for $\gmgchi$ remains approximately satisfied.

\section{The FLRW limit}
\label{sec4b}

From the perturbed LTB equations above it is not clear what each gauge invariant
variable corresponds to in terms of the more familiar Bardeen
potentials, and so forth, of standard cosmology. Here we derive the
standard FLRW perturbation equations in the GMG formalism, and
re-express the GS and GMG variables in terms of FLRW variables.

In the FLRW limit $\kappa \rightarrow$ constant, and we have that
$a_{\perp}$ and $a_{\parallel} \rightarrow a(t)$. The
master equations in the polar sector (\ref{evo1})-(\ref{evo3}) then become
\ba
\nonumber && \Bigg[\p_\tau^2+2\H\p_\tau
-\sdel^2+\frac{4 (1-\kappa r^2)}{r} \p_r-\frac{2}{r^2}
%
%-(1-\kappa r^2)\p_r^2 \\
% &&~~~~~~~~~~~~~+\frac{(2-\kappa
%   r^2)}{r}\p_r+\frac{(\ell-1)(\ell+2)}{r^2} 
%
\Bigg]\gmgchi=0,\nonumber
\ea
and
\ba
&& \nonumber \left[\p_\tau^2+3\H\p_\tau-2\kappa
\right]\gmgk \\&=& \left[-\H\p_\tau+\frac{(1-\kappa r^2)}{r}\p_r+\frac{(\ell(\ell+1)-2+4\kappa r^2)}{2r^2}
\right]\gmgchi,\nonumber
\ea
and
\be
\nonumber
\left[\p_\tau+2\H\right]\gmgpsi=\sqrt{1-\kappa r^2}\p_r\gmgchi,
\ee
while in the axial sector we have
\ba
&&\Bigg[\p_\tau^2+6\H\p_\tau-\sdel^2-\frac{4 (1-\kappa r^2)}{r}
 \left(\p_r+\frac{3}{2r}\right) +6\H^2\Bigg]\gmgPi
\nonumber\\
&&~~~~~~= 16\pi \rho\frac{\sqrt{1-\kappa r^2}}{ar^2}\p_r \gmgbeta,
\ea
and
\be
\p_\tau\gmgbeta=0,
\ee
where $d \tau\equiv dt/a$ is conformal time, and $\H \equiv
a_{,\tau}/a$ is the conformal Hubble rate which obeys the Friedmann and Raychaudhuri  equations
\be
\H^2=\frac{8\pi a^2\rho}{3}-\kappa
~~~~\text{and} ~~~~
\p_\tau\H=-\frac{1}{2}\left(\H^2+\kappa\right).
\ee
Throughout this section $\sdel^2$ will always refer to the Laplacian
acting on a 3-scalar, so that 
\be
\nonumber
\sdel^2=(1-\kappa r^2)\p_r^2+\frac{(2-3\kappa
  r^2)}{r}\p_r-\frac{\ell(\ell+1)}{r^2}.
\ee

It is evident from the equations above that $\chi$ is a
gravitational wave, as its characteristics are null.
However, $\gmgchi$ can also be seen to act as a source for $\gmgk$
even though the homogeneous part of the evolution equation looks like
very similar to the Bardeen 
equation. Similarly, $\gmgpsi$ looks like it almost obeys the vector decay
law, but is also coupled to gravitational waves through $\gmgchi$. The fact that
these variables do not decouple in the FLRW limit means that their
interpretation in terms of FLRW gauge invariants will not be
straightforward.

Let us now consider the perturbed FLRW line-element in the
longitudinal, or conformal Newtonian, gauge:
\bea
\label{FRWcn}
&&\d s^2=-a^2(1+2\Phi)\d\tau^2 -2a^2V_i\d\tau\d x^i 
\\ \nonumber && ~~~~~~~~~+a^2[(1-2\Psi)\gamma_{ij}+h_{ij}]\d x^i\d x^j,
\eea
where $a=a(\tau)$ is the scale factor, and
$\gamma_{ij}=\mathrm{diag}[(1-\kappa r^2)^{-1},r^2\gamma_{ab}]$ is the
background spatial metric in spherical coordinates of curvature
$\kappa$. $\sdel_i$ is the covariant derivative with respect to $\gamma_{ij}$.
The metric (\ref{FRWcn}) is split in the standard way into two 3-scalars ($\Phi,
\Psi$), a 3-vector ($V_i$) and a 3-tensor ($h_{ij}$), where
the `3' is not usually emphasised.  All of these quantities are gauge invariant variables (see
Appendix \ref{FRWperts}).  The vector $V_i$ is divergence-free, and
the tensor $h_{ij}$ is divergence and trace-free. Note that the
coordinates used in the metric in the longitudinal gauge, given by
Eq.~(\ref{FRWcn}), and the metric in the RW gauge, Eqs. (\ref{gpolar})
and (\ref{gaxial}), are not the same.

In Appendix \ref{FRWperts} we show that, in the FLRW limit, the LTB gauge invariants
($\gmgk$, $\gmgpsi$, $\gmgchi$, $\gmgeta$, $\gmgPi$ and $\gmgbeta$) can be written in terms of the usual
FLRW invariants ($\Phi$, $\Psi$, $V_i$ and $h_{ij}$) in the following
way:
\begin{widetext}
Polar:
\ba
\gmgk&=& -2\Psi-2\H V -2\frac{(1-\kappa r^2)}{r}h_r +\frac{1}{r^2}h^\T
+\left[-\H \p_\tau  +\frac{(1-\kappa r^2)}{r} \p_r
 +\frac{\ell(\ell+1)-4(1-\kappa r^2)}{2r^2}\right]h^\TF ,\\  
\gmgpsi&=&\sqrt{1-\kappa r^2} \left\{V_r-\p_r V +\p_\tau h_r -\left(\p_r-\frac{1}{r}\right)\p_\tau h^\TF  \right\},\\
\gmgchi&=& (1-\kappa r^2) h_{rr}+2\left[-(1-\kappa r^2)\p_r+\frac{1}{r} \right]h_r 
-\frac{1}{r^2}h^\T\nonumber\\&& +\left[(1-\kappa r^2)\p_r^2-\frac{(3-2\kappa r^2)}{r}\p_r -\frac{\ell(\ell+1)-8+4\kappa r^2}{2r^2}\right]h^\TF,\\
\gmgeta&=& \Phi-\Psi -\left(\p_\tau +2\H \right)V +\frac{1}{2}(1-\kappa
r^2) h_{rr} +\left[-(1-\kappa r^2)\p_r+\kappa r\right]h_r
\nonumber\\&&+\frac{1}{2}\left[-\p_\tau^2+(1-\kappa r^2)\p_r^2
 -2\H\p_\tau -\frac{(2-\kappa r^2)}{r}\p_r +\frac{2}{r^2} 
\right]h^\TF.
\ea

\end{widetext}

Axial:
\ba
\Pi&=&  \frac{\sqrt{1-\kappa r^2}}{a^2r^2}\left[\left(\p_r-\frac{2}{r}\right)\bar V+\p_\tau\bar h_r\right]              \\
16\pi\rho a \gmgbeta&=& \left[-\sdel^2+\frac{(2-4\kappa r^2)}{r}\p_r -4\kappa\right]\bar V\label{v-bar}
%-\left[(1-\kappa r^2)\p_r+\frac{2-3\kappa r^2}{r} \right]\p_\tau \bar h_r +\frac{(\ell-1)(\ell+2)}{2r^2}\p_\tau\bar h
\ea
%\end{widetext}
where, in order to compare with the GMG formalism, we have split the
3-vector $V_i$ and the 3-tensor $h_{ij}$ into their radial and
angular parts, and then these into their polar and axial components, as described in Appendix \ref{FRWperts}.
This makes explicit the mixing of SVT modes in the LTB gauge
invariants. We have substituted Eq.~(\ref{divh3}) to remove the tensor
contribution to $\gmgbeta$ in Eq.~(\ref{v-bar}).

\subsection{Polar Perturbations}

As we are dealing with FLRW perturbations we may separate the scalar,
vector and tensor parts of these equations, as we know they evolve
independently.

\subsubsection{Scalars} 

From $\gmgeta=0$ we have $\Phi=\Psi$, as is usual in a dust dominated
FLRW cosmology. From the $\gmgk$-equation we then find
\be
\p_\tau^2\Psi+3\H\p_\tau\Psi-2\kappa\Psi=0,
\ee
which is the usual Bardeen equation for the Newtonian potential. The
$\gmgpsi$- and $\gmgchi$-equations contain no scalar modes. We then
note that the scalar part of the gauge invariant matter perturbations can be written
\bea
\label{wscalar}
4\pi a^2\rho\,\gmgomega &=& \sdel^2\Psi-3\H\p_\tau\Psi-3(\H^2-\kappa)\Psi,\\
4\pi a\rho\,\gmgalpha&=&-(\p_\tau+\H)\Psi,\\
4\pi a^2\rho\,\gmggamma&=&-\sqrt{1-\kappa r^2}(\p_\tau+\H)\p_r\Psi.
\eea
It can be seen from (\ref{wscalar}) that the scalar part of the energy
density perturbation is just the gauge invariant density fluctuation
$\delta\rho^\GI\equiv\delta\rho+\p_\tau\rho (B-\p_\tau
E)$~\cite{MFB}.  However, we will see below that we cannot simply
identify $\gmgomega =\delta\rho^\GI/\rho$, as $\gmgomega$ has non-zero
vector and tensor parts.

\subsubsection{Vectors}

From $\gmgeta=0$, and the evolution equation for $\gmgpsi$, we find
that 
\be
\p_\tau V=-2\H V
~~~~\text{and}~~~~
\p_\tau V_r=-2\H V_r. 
\ee
Thus, these two equations give the usual $a^{-2}$ law for vector modes. The
vector part of the evolution equation for $\gmgk$ automatically follows from these two, while the
$\gmgchi$-equation contains no vectors. The vector parts of the gauge invariant
matter perturbations are then given by
\ba
\gmgomega&=&3\H V\\
16\pi a\rho\,\gmgalpha&=& \left[(1-\kappa r^2)\p_r-\kappa r\right]V_r
\\\nonumber &&-\left[(1-\kappa r^2)\p_r^2-\kappa r\p_r-2(3\H^2+\kappa)\right]V,\\
16\pi a^2\rho\,\gmggamma&=&\sqrt{1-\kappa
 r^2}\Big\{\left[\frac{\ell(\ell+1)}{r^2}+3\H^2-\kappa \right]V_r
\\\nonumber &&+\left[\frac{\ell(\ell+1)}{r^2}-3(\H^2+\kappa) \right]\p_r V   
\Big\}.
\ea 
It is a curiosity that the vector part of $\gmgomega=3\H V$ is
non-zero, so that while $\gmgomega$ looks like it is just a gauge invariant density
perturbation, it actually contains vectors.

\subsubsection{Tensors}

The GMG formalism is rather less tidy when trying to describe 3-tensor
modes. The equation which looks like it should be describing
gravitational waves is that for $\gmgchi$. But while this equation consists only of
tensor modes, it is in a very ugly form when evaluated in terms of the FLRW perturbed metric, as
the $\gmgchi$ equation contains fourth derivatives of $h^\TF$. In
fact, as all four GMG variables contain 3-tensors, all four field
equations contain tensor modes, and showing that they reduce to the
usual FLRW wave equation for tensors is non-trivial. The four equations form a set of coupled PDEs for the
variables $h_{rr}, h_r, h^\T, h^\TF$, the simplest of which may be
read off from $\gmgeta=0$ above. Recall, however, that there is only
one degree of freedom in the polar part of $h_{ij}$, with the trace
and divergence free conditions removing the other three. If we choose
this degree of freedom to be $h^\T$ then it can be shown from the GMG equations for
$\gmgeta,\gmgk$ and $\gmgpsi$, when combined with the conditions given by
Eqs. (\ref{h-tf}), (\ref{divh1}) and (\ref{divh2}), reduce to the
single master equation 
\be
[\p_\tau^2+2\H\p_\tau-\sdel^2]h^\T=0,
\ee
which is just the trace of the usual FLRW wave equation.
The $\gmgchi$ equation then follows automatically. 

The tensor parts of the variables associated with the fluid can then
be written 
\ba
\gmgomega&=&\frac{3}{2}\H \p_\tau h^\TF,\label{om_tens}\\ 
\gmgalpha&=&\frac{1}{2}a\p_\tau h^\TF,\label{alp_tens}\\
\gmggamma&=&\frac{1}{2}\sqrt{1-\kappa r^2}\left(\p_\tau h_r+\frac{1}{r}\p_\tau h^\TF\right).
\ea 

\subsection{Axial Perturbations}

There are no scalar modes in the axial sector.

For vector modes it can be seen from $\p_\tau\gmgbeta=0$, and the
radial equation of (\ref{k-axial}), that the usual vector decay law,
$\p_\tau\bar V=-2\H\bar V$, is obeyed. The $\gmgPi$-equation is then automatically satisfied. 

Combining the radial
equation of (\ref{k-axial}) and Eq.~(\ref{divh3}), we find the wave equation
\ba
[\p_\tau^2+2\H\p_\tau-\sdel^2+2\kappa r\p_r+3\kappa]\bar h_r=0,
\ea
where the extra curvature terms are due to $\sdel^2$ acting here on a
scalar that is actually part of a 3-tensor. The equation for $\gmgPi$ then
follows.

\subsection{SVT Variables}
\label{NV}

We have found that what appear as gauge invariant fluid perturbations in
the GMG formalism ~-- namely $\gmgomega$, $\gmgalpha$ and $\gmggamma$~-- are not exclusively
fluid modes, as they are usually understood in FLRW cosmology, as they
can be excited by tensors (FLRW models need a
tensor part in the anisotropic stress for fluid perturbations to
couple to tensor modes, which we have not included here). It is also
the case that the GMG metric perturbations do not correspond in a straightforward way to
the SVT decomposition we are familiar with from FLRW cosmology. For
example, it is clear that $\gmgchi$ represents gravitational waves, and in the
FLRW limit it is a pure tensor mode: but what about scalars and vectors?
Can we identify combinations of gauge invariant GMG variables which will represent purely
scalar or vector modes in the FLRW limit, and so be useful physical
variables in LTB cosmology?

In order to find these, let us first note that the combination 
\be
\lambda\equiv 8\pi a^2\H^{-1}\rho
\left[\gmgomega-3\frac{\H}{a}\gmgalpha\right]
%=8\pi a_\perp H_\perp^{-1}\rho \left[\gmgomega-3H_\perp\gmgalpha\right] 
\ee
does not contain any tensor modes in the FLRW limit (as can be verified
from Eqs.~(\ref{om_tens}) and~(\ref{alp_tens})). Also, note that we
can construct from $\gmgpsi$ and $\gmgchi$ a quantity that
contains only vector modes:
\ba
\nonumber
\xi&\equiv&\frac{3r}{2\sqrt{1-\kappa r^2}}[\sdel^2+3\kappa]{r^{-1}\gmgpsi}+\frac{3}{2r^{2}}\p_r(r^2\p_\tau\gmgchi) \\
%&=&\frac{3}{2W}[\sdel^2+3\kappa]\frac{\psi}{a_\perp
%r}+\frac{3a_{\parallel}}{2r^3W}\left(r^2\dot\gmgchi\right)'\\ 
%&=& \frac{3}{2}\left[(1-\kappa r^2)\p_r^2-3\kappa
%  r\p_r-\frac{\ell(\ell+1)}{r^2}+3\kappa\right]\left(V_r-\p_r
%V\right)\\
&=& \frac{3}{2}\left[ \sdel^2+3 \kappa -\frac{2}{r} \p_r  \right]\left(V_r-\p_r V\right).
\ea
From $\lambda$ and $\xi$ we can now construct a variable that is a
function of scalar modes only:
\ba
\nonumber
\zeta&\equiv&a^{-2}\sdel^2\lambda + a^{-2}\left[(1-\kappa r^2)\p_r+\frac{2-3\kappa r^2}{r}\right]\xi\\
&=&\frac{2}{a^2\H}\left[\sdel^2+3\kappa\right]\sdel^2\Psi.
\ea
This is related to the gauge invariant density fluctuation $\delta\rho^\GI$ and the
curvature perturbation, 
$\mathcal{R}=\Psi-\H(\H\Phi+\p_\tau\Psi)/(\p_\tau\H-\H^2-\kappa)$~\cite{Malik&Wands}, by
\be
\frac{\H\zeta}{8\pi \rho}=\sdel^2\left[\frac{\delta\rho^\GI}{\rho}+3\mathcal{R}-3\Psi\right].
\ee

For the axial perturbations we already see that $\gmgbeta$ is a pure
vector mode, but that $\Pi$ is an awkward mixture of vectors and
tensors. Defining the variable 
\ba
\Upsilon&\equiv&a^{-2}\Bigg[\sdel^2
+\frac{4 (1-\kappa r^2)}{r}\p_r
+\frac{2 (3-4\kappa r^2)}{r^2} \Bigg]\gmgPi \nonumber \\&&
~~~~~~~~~~~~~ +16\pi \rho\frac{\sqrt{1-\kappa r^2}}{a^3 r^2}\p_r\gmgbeta \\
&=& \frac{\sqrt{1-\kappa r^2}}{a^4r^2}
\Bigg[\sdel^2-2 \kappa r \p_r-\kappa
\Bigg]\p_\eta\bar h_r
\ea
we are able to define a pure tensor mode using the GMG gauge invariant
variables.

Using these new variables, we can now generalise to the inhomogeneous
case. The resulting functions will have the useful feature that
they reduce to pure scalar or vector modes in the FLRW limit.  An example set of
variables is given by Eqs. (\ref{te1})-(\ref{sc1}), in section \ref{newvars}.
These reduce to the quantities above in the FLRW limit. The variables
$\hat{\zeta}$, $\hat{\xi}$, $\hat{\gmgbeta}$, $\hat{\gmgchi}$ and $\hat{\Upsilon}$, are then useful gauge invariant variables in
the LTB perturbation equations as they encode generalised scalar, vector
and tensor perturbations, respectively, as they are normally referred to in
cosmology.

\section{Discussion}
\label{sec5}

We have presented here a full system of master equations that represent the general perturbations to LTB
space-times, in terms of the gauge invariant variables $\gmgk,
\gmgpsi,\gmgchi,\gmgPi,\gmgbeta$ (and $\gmgeta$, strictly
speaking). This formalism can now be used to investigate, in a fully
consistent fashion, the growth of linear structure in LTB models.  As
such, predictions for phenomena such as baryon acoustic oscillations,
the integrated Sachs-Wolfe effect, and the observed matter power spectrum can
now be made, and used to compare with astrophysical data.  This will
allow us to establish what deviations should be expected from the usual
FLRW predictions, and whether or not such deviations can be observed
in our Universe.

However, while $\gmgk, \gmgpsi,\gmgchi,\gmgPi,\gmgbeta$ and $\gmgeta$
appear completely naturally, and produce an elegant system of equations in the form of an
initial value formulation, they do not reduce to anything intuitive
in the FLRW limit. Instead, they are a cumbersome mixture of scalar,
vector and tensor perturbations~-- a situation made worse still by the
additional couplings that exist in the evolution equations in the more
general, inhomogeneous case.

To address these problems we have defined a new set of variables in
the polar sector, $\hat{\zeta}$, $\hat{\xi}$ and $\hat{\gmgchi}$, that can be thought of as
generalised scalar, vector and tensor perturbations, respectively.
Similarly,  we propose new variables in the axial sector, $\hat{\gmgbeta}$ and
$\hat{\Upsilon}$, that correspond to generalised vector and
tensor modes.  In the FLRW
limit these new variables become pure scalars, vectors and tensors.
We expect these new variables to prove useful in future studies where they can be used to set the initial
conditions for the evolution of perturbations.  For example, in an LTB
model with homogeneous bang time, $t_0(r)=$constant, the surfaces of
constant $t$ will be almost homogeneous at early times, and will then be
well approximated by an FLRW description.  The new variables can then be
used to match the FLRW initial conditions onto the LTB evolution
equations in a straightforward way.  This will be considered further in future studies.

\appendix

\section{The Gerlach-Sengupta formalism}
\label{A}

The general form of perturbations to the metric and stress-energy
tensors can be written, in terms of the harmonic functions outlined in
Section \ref{sec2}, as $g_{\mu \nu} \rightarrow g_{\mu \nu} + \Delta
g_{\mu \nu}$ and $T_{\mu \nu} \rightarrow T_{\mu \nu} + \Delta
T_{\mu \nu}$, where \cite{gersen}
\begin{align}
\Delta g_{\mu \nu} 
&\equiv
\left( \begin{array}{cc}
0 & h_A^{\text{axial}} \bar Y_a\\
h_A^{\text{axial}} \bar Y_a \; & h \; \bar Y_{ab} \end{array} \right)\\
\Delta T_{\mu \nu} 
&\equiv
\left( \begin{array}{cc}
0 & \Delta t_A^{\text{axial}} \bar Y_a\\
\Delta t_A^{\text{axial}} \bar Y_a \; & \Delta t^{(1)} \; \bar Y_{ab} \end{array} \right),
\end{align}
for axial perturbations, and
\begin{align}
\Delta g_{\mu \nu} 
&\equiv
\left( \begin{array}{cc}
h_{AB} Y & h_A^{\text{polar}} Y_{a}\\
h_A^{\text{polar}} Y_{a} \; \; & a_{\perp}^2 r^2 (K Y \gamma_{ab}+G Y_{:ab}) \end{array} \right)\\
\Delta T_{\mu \nu} 
&\equiv
\left( \begin{array}{cc}
\Delta t_{AB} Y & \Delta t_A^{\text{polar}} Y_a\\
\Delta t_A^{\text{polar}} Y_a \; & \Delta t^{(2)} Y_{ab} +  a_{\perp}^2 r^2 \Delta t^{(3)} Y
\gamma_{ab} \end{array} \right),
\end{align}
for polar perturbations. The new variables
$h_A^{\text{axial}}$, $h$, $\Delta t_A^{\text{axial}}$, $\Delta t^{(1)}$,
$h_{AB}$, $h_A^{\text{polar}}$, $K$, $G$, $\Delta t_{AB}$, $\Delta
t_{A}^{\text{polar}}$, $\Delta t^{(2)}$ and $\Delta t^{(3)}$ are all functions
of the two coordinates $x^A$.

From these quantities we can then construct the gauge invariant
variable \cite{gersen}
\be
k_A \equiv h_A^{\text{axial}}- h_{\vert A}+2 h v_A,
\ee
for axial perturbations, and
\begin{align}
k_{AB} &\equiv h_{AB} - p_{A \vert B} - p_{B \vert A}\\
\gmgk &\equiv K - 2 v^A p_A\\
t_{AB} &\equiv \Delta t_{AB} - t_{AB \vert C} p^C - t_{AC}
{p^C}_{\vert B} - t_{BC} {p^C}_{\vert A}\\
t_A &\equiv \Delta t_A^{\text{polar}} - t_{AC} p^C,
\end{align}
for scalar perturbations.  The quantities $\Delta
t_A^{\text{axial}}$, $\Delta t^{(1)}$, $\Delta t^{(2)}$
and $\Delta t^{(3)}$ are already gauge invariant for a dust
dominated universe.  Here a pipe indicates a covariant derivative with
respect to the metric on $M^2$, and we have defined $p_A
\equiv h_A^{\text{polar}}-a_{\perp}^2 r^2 G_{\vert A}/2$, along with GS, for
concision.  All of these expressions are constructed so as to be
invariant under infinitesimal coordinate transformations of the form $x^{\mu}
\rightarrow x^{\mu} + \xi^{\mu}$.

There exists here a useful gauge in which $h=
h_A^{\text{polar}}=G=0$.  This is the Regge-Wheeler gauge \cite{RW}, where the
remaining perturbation variables are all equal to gauge invariant
quantities, as can be seen from the expressions above.

\section{The Gundlach-Mart\'{i}n-Garc\'{i}a formalism}
\label{B}

The background quantities used by GMG can be written in the LTB case,
using our notation, as
\ba
U &=& H_{\perp}\\
V &=& \frac{H_{\perp}}{W}\\
m &=& \frac{M}{2}\\
\mu &=& H_{\parallel}\\
\nu &=& 0.
\ea
GMG then write their perturbed fluid four velocity as $u_{\mu} \rightarrow u_{\mu}
+\Delta u_{\mu}$, where
\be
\Delta u_{\mu} = (0, \gmgbeta \; \bar Y_a)
\ee
for axial perturbations, and
\be
\Delta u_{\mu} = \left[ \left(\tilde{\gmggamma} n_A+\frac{1}{2}h_{AB} u^B
\right) Y, \tilde{\gmgalpha} \; Y_a \right]
\ee
for scalar perturbations.  Here $n_A\equiv - \epsilon_{AB} u^B$ is a
unit space-like vector, and $\tilde{\gmgalpha}$, $\gmgbeta$ and
$\tilde{\gmggamma}$ are all functions of $x^A$.  Density perturbations
are then parametrised by $\rho \rightarrow \rho + \tilde{\gmgomega} Y
\rho$, and the following gauge invariant quantities can be
constructed:
\begin{align}
\gmgalpha &\equiv \tilde{\gmgalpha}- p^B u_B\\
\gmggamma &\equiv \tilde{\gmggamma} - n^A u_{A \vert B} p^B + \frac{1}{2}
n^A u^B (p_{A \vert B} - p_{B \vert A})\\
\gmgomega &\equiv \tilde{\gmgomega} - p^A (\ln \rho)_{\vert A}
\end{align}
with $\gmgbeta$ already gauge invariant.  Again, the Regge-Wheeler gauge
can be seen to have a special significance.  The gauge invariants
above can be related to those in Appendix \ref{A} via
\begin{align}
L_A &= \gmgbeta u_A\\
t_A &= \gmgalpha \rho u_A\\
t_{AB} &= \rho \Big[ \gmggamma (u_A n_B + n_A u_B) +\gmgomega u_A u_B \\
&\qquad \qquad \qquad + \frac{1}{2} (k_{AC} u_B + u_A k_{BC}) \Big]
\end{align}
together with $L=t^{(2)}=t^{(3)}=0$.  GMG then continue to decompose
$k_{AB}$ into the `fluid frame' via
\begin{align}
k_{AB} \equiv \gmgeta (n_A n_B - u_A u_B) &+ \gmgphi (n_A n_B + u_A u_B) \\
&+ \gmgpsi (u_A n_B+ n_A u_B),
\end{align}
and to define the new variable
\be
\gmgchi \equiv \gmgphi-\gmgk+\gmgeta.
\ee
In these variables the perturbation equations take a particularly
simple form \cite{gmg}.

\begin{widetext}

\section{Polar $\ell =0,1$ perturbations}
\label{C}

For polar perturbations with $\ell =0,1$ it is no longer the case that the
field equations give $\gmgeta=0$.  The general system of equations
given in Sec~\ref{sec3a} now reads:

\noindent For $\ell\geq1$:

\bea
\label{evo1-eta}
&& -\ddot{\gmgchi} + \gmgchi^{\prime \prime} + 2 (H_{\parallel} -H_{\perp})
\gmgpsi^{\prime} - 2\gmgeta^{\prime \prime}\\ \nonumber
&=& -2 \left[ 8 \pi \rho -\frac{3M}{a_{\perp}^3}-2 H_{\perp} (H_{\parallel}-H_{\perp})
\right] (\gmgchi + \gmgk) 
+\frac{(\ell -1)(\ell +2)}{a_{\perp}^2 r^2} \gmgchi +3 H_{\parallel}
\dot{\gmgchi}
+ 4 (H_{\parallel} -H_{\perp}) \dot{\gmgk} +2 W \gmgchi^{\prime} \\
\nonumber && -2  \left[
H_{\parallel}^{\prime}-2(H_{\parallel}-H_{\perp}) W \right] \gmgpsi  - 2
(H_{\parallel}-H_{\perp}) \dot{\gmgeta}    - 6W \gmgeta^{\prime}   - \left[
\frac{\ell (\ell +1)+8}{a_{\perp}^2 r^2} +8 H_{\parallel} H_{\perp} + 4
H_{\perp}^2 -16 W^2  -32 \pi \rho \right]\gmgeta,
\eea
\bea
\label{evo2-eta}
- \ddot{\gmgk}  &=& 4 H_{\perp} \dot{\gmgk} +\left[ 2 W^2 -\frac{\ell (\ell +1)+2}{2
  a_{\perp}^2 r^2} \right] (\gmgchi-2 \gmgeta) -2 W
(H_{\parallel} -H_{\perp}) \gmgpsi \\ \nonumber && +H_{\perp} (\dot{\gmgchi}-2 \dot{\gmgeta}) - W (\gmgchi^{\prime}-2
\gmgeta^{\prime})-2 W^2 \gmgeta -2 \left( \frac{1}{a_{\perp}^2 r^2}-W^2 \right) \gmgk,
\eea
\be
\label{evo3-eta}
-\dot{\gmgpsi} = 2 H_{\parallel} \gmgpsi + \gmgchi^{\prime}+2 W \gmgeta -2 \gmgeta^{\prime} .
\ee
\be
\label{con3-eta}
8 \pi \rho \gmgalpha = \frac{\gmgpsi^{\prime}}{2} + H_{\parallel} (\gmgchi+\gmgk)+ \frac{\dot{\gmgchi}}{2}
+\dot{\gmgk} -(H_{\parallel}+H_{\perp}) \gmgeta.
\ee
\be
\label{mat1-eta}
\dot{\gmgalpha}= \frac{\gmgchi}{2} +\frac{\gmgk}{2}-\gmgeta,
\ee

\noindent and for $\ell\geq0$:

\bea
\label{con1-eta}
8 \pi \rho \gmggamma &=& ( \dot{\gmgk} )^{\prime} - W \dot{\gmgchi}
+H_{\perp} \gmgchi^{\prime} - (H_{\parallel} -2 H_{\perp})
\gmgk^{\prime} - 2 H_{\perp} \gmgeta^{\prime}  + \left[ \frac{\ell  (\ell +1)+2}{2 a_{\perp}^2 r^2} +H_{\perp}^2 +2 H_{\perp} H_{\parallel} -
W^2 -4 \pi \rho \right] \gmgpsi,
\eea
\bea
\label{con2-eta}
8 \pi \rho \gmgomega &=&  \left[ \frac{\ell (\ell +1)}{a_{\perp}^2r^2} +2 H_{\perp}^2 +4 H_{\parallel} H_{\perp} -8 \pi \rho \right]
(\gmgchi +\gmgk)+2 H_{\perp} \gmgpsi^{\prime}  - \frac{(\ell -1)(\ell +2)}{2 a_{\perp}^2r^2} \gmgchi \\ \nonumber && +2 (H_{\parallel}+H_{\perp}) W
\gmgpsi + H_{\perp} \dot{\gmgchi} -\gmgk^{\prime \prime} +(H_{\parallel}+2 H_{\perp}) \dot{\gmgk} +W
\gmgchi^{\prime} - 2 W \gmgk^{\prime}-2 H_{\perp} (H_{\perp} +2
H_{\parallel}) \gmgeta,
\eea
\be
\label{mat2-eta}
\dot{\gmggamma} = \frac{\gmgk^{\prime}}{2} - H_{\parallel} \gmggamma -
\frac{H_{\parallel}}{2} \gmgpsi -W \gmgeta,
\ee
\bea
\label{mat3-eta}
\dot{\gmgomega}+\left( \gmggamma +\frac{\gmgpsi}{2}
\right)^{\prime} =
\frac{\ell  (\ell +1)}{a_{\perp}^2r^2} \gmgalpha 
-\left(
\gmggamma +\frac{\gmgpsi}{2} \right) \frac{\rho^{\prime}}{\rho}  
-\frac{\dot{\gmgchi}}{2} -\frac{3
\dot{\gmgk}}{2} -2 W \left( \gmggamma +\frac{\gmgpsi}{2} \right).
\eea

\subsection{$\ell=1$}

For $\ell=1$ there is an additional gauge freedom which can
be used to eliminate one of the perturbation variables.  The obvious
choice may be to use this to set $\gmgeta =0$, but GMG showed this to
lead to some ambiguity.  We therefore follow them in using the less
ambiguous choice $\gmgk =0$.  The evolution and conservations
equations above can then be combined to give
\ba
\label{l1a}
W \gmgeta^{\prime} - H_{\perp} \dot{\gmgeta} &=& 4 \pi \rho \gmgomega -16 \pi \rho H_{\perp} \gmgalpha - \left(
\frac{2}{a_{\perp}^2 r^2}-3 W^2 +H_{\perp}^2 \right) \gmgeta - \left[
W^2+H_{\perp}^2-4 \pi \rho \right] \gmgchi -2 H_{\perp}W \gmgpsi,\\
\label{l1b}
W \gmgchi^{\prime} - H_{\perp} \dot{\gmgchi} &=& 8 \pi \rho \gmgomega - 32 \pi \rho H_{\perp} \gmgalpha - \left[
\frac{2}{a_{\perp}^2 r^2} +2 H_{\perp}^2 -8 \pi \rho \right]\gmgchi - 2
(H_{\parallel}+H_{\perp}) W \gmgpsi -2 H_{\perp}^2 \gmgeta,
\\
\label{l1c}
W \gmgpsi^{\prime} - H_{\perp} \dot{\gmgpsi} &=& 8 \pi \rho (\gmggamma +2 W \gmgalpha) + 2 H_{\parallel} W
(\gmgeta-\gmgchi)  + 4
H_{\perp} W \gmgeta - \left[ H_{\perp}^2-W^2+\frac{2}{a_{\perp}^2 r^2} -4\pi \rho
\right] \gmgpsi.
\ea
These are the LTB equivalents of (GMGA12), (GMGA13) and (GMGA14) from
\cite{gmg}.  These three equations can be solved, along with the three conservation equations
(\ref{mat1})-(\ref{mat3}), with $\gmgk =0$, for the six remaining
perturbation variables, $\gmgeta$, $\gmgchi$, $\gmgpsi$, $\gmgalpha$,
$\gmggamma$ and $\gmgomega$.

\subsection{$\ell =0$}
\label{D}

Polar perturbations with $\ell =0$ are spherical.  As such one
could conceive of absorbing such perturbations into the background
parameters, which themselves describe the most general spherically
symmetric dust dominated solution of Einstein's equations. Despite
this being the case, it still seems worth considering $\ell =0$
perturbations in a spherical background, as one can then separate them
out more easily from a smooth background.

In this case we again have a non-zero $\gmgeta$,
but now have sufficient gauge freedom to set $\gmgk=0$ and
\be
\gmgpsi=\frac{2 H_{\perp} W}{H_{\perp}^2+W^2} (\gmgeta- \gmgchi),
\ee
as done by GMG.  The constraint equations then give
\bea
\label{l0a}
W\gmgeta^{\prime} - H_{\perp} \dot{\gmgeta} &=& 4 \pi \rho \left( \gmgchi +\frac{2 H_{\perp}^2}{(W^2-H_{\perp}^2)} \gmgeta
\right) + 16 \pi \rho \frac{H_{\perp}W}{(W^2-H_{\perp}^2)} \gmggamma + 4
\pi \rho \frac{(W^2+H_{\perp}^2)}{(W^2-H_{\perp}^2)} \gmgomega,
\eea
and
\bea
\label{l0b}
W\gmgchi^{\prime} - H_{\perp} \dot{\gmgchi} &=& \frac{4 H_{\perp} W}{(H_{\perp}^2+W^2)} \left[ H_{\perp} W+4 \pi
\rho \frac{H_{\perp}W}{(W^2-H_{\perp}^2)} \right] (\gmgchi-\gmgeta) +16
\pi \rho \frac{H_{\perp}W}{(W^2-H_{\perp}^2)} \gmggamma \\ \nonumber &&  +8 \pi \rho
\frac{(W^2+H_{\perp}^2)}{(W^2-H_{\perp}^2)} \gmgomega  
+ \left( 8\pi
\rho - \frac{1}{a_{\perp}^2r^2} \right) \left( \gmgchi +\frac{2
H_{\perp}^2}{(W^2-H_{\perp}^2)} \gmgeta \right).
\eea
These are the LTB versions of (GMGA18) and (GMGA19).
The two remaining conservation equations are then
\bea
\label{l0c}
\dot{\gmggamma} &=& - \left[ H_{\parallel}+4 \pi \rho
\frac{H_{\perp}}{(W^2-H_{\perp}^2)} \right] \gmggamma -\frac{W}{2a_{\perp}^2r^2
(W^2-H_{\perp}^2)} \gmgeta \\ \nonumber && + (\gmgeta-\gmgchi) \Bigg[
\frac{W(W^2-H_{\perp}^2)}{2 (W^2+H_{\perp}^2)} - \frac{H_{\parallel}
H_{\perp} W}{(W^2+H_{\perp}^2)}  +4 \pi \rho \frac{W
  H_{\perp}^2}{(H_{\perp}^4-W^4)} \Bigg],
\eea
and
\bea
\dot{\gmgomega} + \gmggamma^{\prime} &=&  \frac{4 \pi \rho H_{\perp}}{(W^2-H_{\perp}^2)} \gmgomega 
-\frac{H_{\perp}}{(W^2-H_{\perp}^2)} \left( \frac{1}{2 a_{\perp}^2r^2}-4 \pi \rho
\right) \gmgeta 
- \left[
\frac{\rho^{\prime}}{\rho} +2W-4 \pi \rho
\frac{W}{(W^2-H_{\perp}^2)}  \right] \gmggamma
\\ \nonumber && -(\gmgeta - \gmgchi) \Bigg[ \frac{H_{\perp} W}{(W^2 +H_{\perp}^2)}
  \frac{\rho^{\prime}}{\rho} +H_{\perp}+H_{\parallel} +
  \frac{H_{\perp} (W^2-H_{\perp}^2)}{2 (W^2+H_{\perp}^2)} + 4 \pi
  \rho \frac{H_{\perp} W^2}{(W^4-H_{\perp}^4)} \Bigg].
\eea
These equations are the LTB versions of (GMGA15) and (GMGA16).  We now
have four equations to solve for the four
remaining variables $\gmgeta$, $\gmgchi$, $\gmggamma$ and $\gmgomega$.

\end{widetext}

\section{Matching LTB and FLRW gauge invariants}
\label{FRWperts}

Here we derive the standard FLRW perturbation equations in the GMG
formalism.  Unfortunately, the RW gauge is not well
adapted to our usual description of FLRW perturbations. We must
therefore write both the GMG variables and the perturbed FLRW metric in a general gauge.

In general coordinates an arbitrary perturbation of FLRW can be written
\bea
&&\d s^2=-a^2(1+2\phi)\d\tau^2 +2a^2(\sdel_iB-S_i)\d\tau\d x^i 
\\&&~~~ \nonumber +a^2 \big[ (1-2\psi)\gamma_{ij}+2\sdel_i\sdel_j
 E +2\sdel_iF_j+h_{ij} \big]\d x^i\d x^j,
\eea
where $a=a(\tau)$, $\gamma_{ij}$ is the spatial metric, and $\sdel_i$
is the covariant derivative with respect to $\gamma_{ij}$.  The
vectors here are divergence free, $\sdel_i S^i=0$, and the tensors are divergence and trace-free. Gauge
invariant metric perturbations are given by 
\ba
\Phi&=&\phi+\mathcal{H}(B-\p_\tau E)+\p_\tau(B-\p_\tau E),\nonumber\\ 
\Psi&=&\psi-\mathcal{H}(B-\p_\tau E),\nonumber\\
V_i&=& S_i+\p_\tau F_i,\label{GIFLRW}
\ea
and the perturbations $h_{ij}$ are already gauge-invariant.

To compare with the GMG formalism, we expand this 1+3 split into a
1+1+2 split in spherical coordinates, peeling off the radial parts of
each variable. A 3-vector, such as $S_i$, then splits into a
scalar, $S_r$, and a 2-vector, $S_a$. These can then be decomposed into
spherical harmonics as 
\ba
S_r&=&\sum_{\ell m} S_r^\lm Y^\lm=S_rY,\nonumber\\
S_a&=&\sum_{\ell m} S^\lm Y^\lm_{:a}+\bar{S}^\lm
\epsilon_{a}^{~b}Y^\lm_{:b},\nonumber\\
&=& S Y_a + \bar{S} \bar Y_a,
\ea
where $S$ and $\bar S$ are the polar and axial parts of $S$,
respectively. The divergence-free property of the 3-vector, $\sdel_i
S^i=0$, then gives us that
\be
(1-\kappa r^2)\p_r S_r +\frac{(2-3\kappa r^2)}{r}S_r-\frac{\ell(\ell+1)}{r^2}S=0.
\ee
This reduces the degrees of freedom in the 3-vector to two: one polar and one axial.

For a 3-tensor, the split into radial and angular parts is a bit more messy. The tensor 
\be
h_{ij}=\left(
\begin{array}{c|c}
h_{rr}   & h_{ra}  \\ \hline  
h_{ra}  & h_{ab}  \\
\end{array}
\right)
\ee 
splits into a 2-scalar $h_{rr}=h_{rr}^\lm Y^\lm$ (spherical
harmonic sum implied), a 2-vector $h_{ra}= h_r^\lm
Y^\lm_{a}+\bar{h}_r^\lm \bar Y^\lm_{a}$ and a remaining
part consisting of a 2-scalar (the trace) and polar and axial
trace-free 2-tensors:
\be
h_{ab}= h^{\T} \gamma_{ab} Y +h^{\TF} Y_{ab} + \bar h \bar Y_{ab}.
\ee
This splits the 3-tensor $h_{ij}$ into its polar ($h_{rr}$, $h_r$, $h^{(T)}$
and $h^{(TF)}$) and axial ($\bar{h}_r$ and $\bar{h}$) parts.

The trace-free, $\gamma^{ij}h_{ij}=0$, and divergence-free, $\sdel^i h_{ji}=0$, conditions give:
\begin{widetext}
\ba
0 &=& (1-\kappa r^2) h_{rr}+\frac{2}{r^2} h^\T ,\label{h-tf}\\
0 &=& (1-\kappa r^2)\partial_r h_{rr}+\frac{(2-4\kappa r^2)}{r}h_{rr}
%\nonumber\\ && 
-\frac{2}{r^3}h^\T-\frac{\ell(\ell+1)}{r^2}h_r ,\label{divh1}\\
0 &=& (1-\kappa r^2)\partial_r h_{r}+\frac{(2-3\kappa r^2)}{r}h_{r} 
%\nonumber\\ &&
+\frac{1}{r^2}h^\T-\frac{(\ell-1)(\ell+2)}{2r^2}h^\TF,\label{divh2}\\
0 &=& (1-\kappa r^2)\partial_r \bar h_{r}+\frac{(2-3\kappa r^2)}{r}\bar
h_{r} 
%\nonumber\\&&
-\frac{(\ell-1)(\ell+2)}{2r^2}\bar h. \label{divh3}
\ea 
Again, the degrees of freedom left after applying these relations is two: one per parity. 

We can now equate the perturbed FLRW with the perturbed LTB metric, in
an arbitrary gauge, to find that the polar perturbations are related by
\ba
a^{-2}h^\GMG_{\tau\tau}&=&-2\phi,\\
a^{-2}h^\GMG_{rr}&=& -\frac{2}{1-\kappa r^2}\psi
+2\left(\p_r-\frac{\kappa r}{1-\kappa r^2}\right)\p_r E
+2\left(\p_r-\frac{\kappa r}{1-\kappa r^2}\right)F_r+h_{rr},\\ 
a^{-2}h^\GMG_{r\tau}&=& -S_r+\p_rB ,\\
a^{-2}h^\GMG_{\tau}&=&-S+B,\\
a^{-2}h^\GMG_r &=& 2\left(\p_r-\frac{1}{r}\right)E+F_r+\left(\p_r-\frac{2}{r}\right)F+h_r,\\
r^{2}G &=& 2E+2F+h^\TF,\\
K&=& -2\psi +2\frac{1-\kappa r^2}{r}\p_r E+2\frac{1-\kappa r^2}{r}F_r +\frac{1}{r^2}h^\T+\frac{\ell(\ell+1)}{2r^2}h^\TF,
\ea
where everything has been decomposed into spherical harmonics, so that
$\phi=\phi^\lm(x^A)$ etc. 
For the axial modes we have
\ba
a^{-2}h^\GMG_\tau&=& -\bar S,\\
a^{-2}h^\GMG_r&=&\left(\p_r-\frac{2}{r}\right)\bar F+\bar h_r, \\
a^{-2}h^\GMG&=&\bar F+\frac{1}{2}\bar h ,
\ea
The gauge invariant GMG variables
can now be calculated
directly, and are given in Section \ref{sec4b}.

\end{widetext}

% ------------------------ ACKNOWLEDGEMENTS ----------------------------------
%\vspace{-10pt}
\section*{Acknowledgements}
%\vspace{-10pt}

We are grateful to Pedro G. Ferreira and Roy Maartens for helpful discussions and comments.
TC acknowledges the support of Jesus College and the BIPAC. CC acknowledges funding from the National Research
Foundation (South Africa) and the University of Cape Town. SF is
funded by the National Astrophysics and Space Science Programme (South
Africa) and the NRF. CC would like to thank  
the Department of Astrophysics at the University of Oxford for hospitality while some of the work presented here was undertaken. 
This visit was partly supported by a joint Royal Society (UK) - NRF grant UID 65329.

%\end{widetext}

% ---------------------- BIBLIOGRAPHY -----------------------------------------
%\vspace{-10pt}

%\end{widetext}

\end{document}